\documentclass[runningheads]{llncs}
\usepackage[T1]{fontenc}
\usepackage{graphicx}
\usepackage{multirow}
\usepackage{amsmath}
\usepackage{amsfonts}
\usepackage{amssymb}
\usepackage{color}
\usepackage{adjustbox}
\usepackage{marvosym}

\def\model{VQ-Seq2Seq}

\begin{document}
\title{Non-Adversarial Learning: Vector-Quantized Common Latent Space for Multi-Sequence MRI}
\titlerunning{Vector Quantized Common Latent Space}
%
%
\author{Luyi Han\inst{1,2}\orcidID{0000-0003-4046-2763} \and
Tao Tan\inst{3,2}\textsuperscript{(\Letter)} \and
Tianyu Zhang\inst{1,2,4} \and
Xin Wang\inst{2,4} \and
Yuan Gao\inst{2,4} \and
Chunyao Lu\inst{1,2} \and
Xinglong Liang\inst{1,2} \and
Haoran Dou\inst{5} \and
Yunzhi Huang\inst{6} \and
Ritse Mann\inst{1,2}
}
%
\authorrunning{Han et al.}
%
\institute{Department of Radiology and Nuclear Medicine, Radboud University Medical Centre, Geert Grooteplein 10, 6525 GA, Nijmegen, The Netherlands \and
Department of Radiology, The Netherlands Cancer Institute, Plesmanlaan 121, 1066 CX, Amsterdam, The Netherlands \and
Faculty of Applied Sciences, Macao Polytechnic University, 999078, Macao Special Administrative Region of China \and
GROW School for Oncology and Developmental Biology, Maastricht University Medical Centre, P. Debyelaan 25, 6202 AZ, Maastricht, The Netherlands \and
Center for Computational Imaging and Simulation Technologies in Biomedicine within the School of Computing at the University of Leeds, LS2 9JT Leeds, UK \and
School of Artificial Intelligence, Nanjing University of Information Science and Technology, Nanjing 210044, China
\\
\email{taotan@mpu.edu.mo}}
\maketitle              
\begin{abstract}
Adversarial learning helps generative models translate MRI from source to target sequence when lacking paired samples. However, implementing MRI synthesis with adversarial learning in clinical settings is challenging due to training instability and mode collapse.
To address this issue, we leverage intermediate sequences to estimate the common latent space among multi-sequence MRI, enabling the reconstruction of distinct sequences from the common latent space.
We propose a generative model that compresses discrete representations of each sequence to estimate the Gaussian distribution of vector-quantized common (VQC) latent space between multiple sequences. Moreover, we improve the latent space consistency with contrastive learning and increase model stability by domain augmentation.
Experiments using BraTS2021 dataset show that our non-adversarial model outperforms other GAN-based methods, and VQC latent space aids our model to achieve (1) anti-interference ability, which can eliminate the effects of noise, bias fields, and artifacts, and (2) solid semantic representation ability, with the potential of one-shot segmentation.
Our code is publicly available~\footnote{\url{https://github.com/fiy2W/mri\_seq2seq}}.

\end{abstract}
\section{Introduction}
Multi-sequence magnetic resonance imaging (MRI) is a commonly used diagnostic tool that provides clinicians with a comprehensive view of tissue characteristics~\cite{chen2013clinical,menze2014multimodal,mann2019breast}.
However, some sequences may be unusable or absent in clinical practice for various reasons~\cite{chartsias2017multimodal}, leading to the need for rescanning or disrupting downstream processes.
To avoid this, deep generative models can be used to synthesize these missing sequences, but require many paired training data to produce high-quality results. In cases lacking paired data among source and target sequence, most studies~\cite{sharma2019missing,jiang2020unified,dalmaz2022resvit} rely on generative adversarial networks (GANs)~\cite{goodfellow2020generative} to minimize the distribution distance between the generated and the target sequence. However, it can also lead to training instability and mode collapse, harming the image quality and structure.

Using intermediate sequences in multi-sequence MRI can make unsupervised generation less challenging.
For example, if we have paired T1-weighted (T1) and T2-weighted (T2) MRI for one population and paired T2 and fluid-attenuated inversion recovery (Flair) MRI for another, we can use T2 to establish the relationship between T1 and Flair without paired samples.
Compared to single-task models~\cite{sharma2019missing,dalmaz2022resvit}, dynamic models~\cite{jiang2020unified,chen2023unsupervised,han2024synthesis} controlled by a prompt branch can integrate multiple generation tasks to utilize intermediate sequences. Han~\textit{et al.}~\cite{han2024synthesis} use a shared encoder to extract structural features from images, which are then rendered to target images with the guidance of a one-hot code. Jiang~\textit{et al.}~\cite{jiang2020unified} disentangle images into structure and style features and reconstruct target images using target styles and source structures.
These methods preserve the structure consistency but ignore the distribution differences between the latent spaces of distinct sequences, hindering the model from learning the mapping of the common latent space to the target sequence.

In this work, we construct a common latent space for multi-sequence MRI so that all sequences can be mapped from it. Specifically, we first utilize VQ-VAE~\cite{van2017neural} to compress images into a discrete latent space, then estimate the distribution of the vector-quantized common (VQC) latent space based on these representations. Finally, we leverage a dynamic model Seq2Seq~\cite{han2024synthesis} to generate arbitrary target sequences from the VQC latent space.
The VQC latent space has three primary advantages: (1) achieving unsupervised synthesis without requiring adversarial learning; (2) preventing input interference, such as noise, artifacts, and field bias; and (3) having reliable semantic representation, which shows the potential of one-shot segmentation.

\section{Methods}
\subsection{Preliminary}
\subsubsection{VQ-VAE}\label{sec:vqvae}
Compared with the continuous latent space of VAE~\cite{kingma2013auto}, the discrete latent space of VQ-VAE captures more structured features while ignoring some irrelevant details, \textit{e.g.,} artifacts.
Given an encoder $\mathbf{E}$ and a decoder $\mathbf{G}$, we can map image $X$ into a continuous latent space $z_e=\mathbf{E}(X)$ with a latent dimension of $D$, while $\mathbf{G}$ can restore $X$ from $z_e$.
Then, using a codebook $e_k$ with the embedding dimension of $K$ to map $z_e$ to the nearest vectors in the codebook.
\begin{equation}
    \label{eq:vq}
    z_q(z_e)=e_k, \quad k=\arg\min_i\left\|z_e-e_i\right\|
\end{equation}
where the vector quantizing process is not differentiable, requiring an improved training loss,
\begin{equation}
    \label{eq:vqvae}
    \mathcal{L}_\text{vqvae} = \left\|X-\mathbf{G}(z_e+\text{sg}\left[z_q-z_e\right])\right\|_2^2+\left\|\text{sg}\left[z_e\right]-z_q\right\|_2^2+\beta\cdot\left\|\text{sg}\left[z_q\right]-z_e\right\|_2^2
\end{equation}
where $\text{sg}\left[\cdot\right]$ indicates a stop-gradient operation, and $\beta=0.25$ ensures that $z_e$ remains in proximity to $z_q$. To simplify the expression, we denote $z_e+\text{sg}\left[z_q-z_e\right]$ as $z_q$ and merge the last two terms of Eq.~\ref{eq:vqvae} as $\mathcal{L}_\text{vq}$ in the following sections.
\subsubsection{Dynamic Model}
Dynamic models~\cite{jiang2020unified,chen2023unsupervised,han2024synthesis} combine different generation tasks in a single model, which can utilize intermediate sequences. With a set of $N$ sequences MRI $\mathcal{X}=\left\{X_i,f_i|i=1,...,N\right\}$, $X_i$ is available if $f_i = 1$, otherwise $f_i = 0$ and the sequence is missing. The process of translating $X_i$ to $X_j$ is,
\begin{equation}
    \label{eq:seq2seq}
    \hat{X}_{i\rightarrow j}=\mathbf{G}(\mathbf{E}(X_i),c_j)
\end{equation}
where $\mathbf{G}$ refers to a dynamic decoder which input with structure feature $\mathbf{E}(X_i)$ and style feature $c_j$. In particular, $c_j$ can be represented as a one-hot encoding for the target sequence~\cite{chen2023unsupervised,han2024synthesis} or a style feature extracted from the target image~\cite{jiang2020unified}.
In this work, we use Seq2Seq~\cite{han2024synthesis} as the baseline because the model is a simple autoencoder, which makes it easy to integrate the VQ module into the model.
\subsection{\model}
\begin{figure}[t]
\centering
    \centerline{\includegraphics[width=0.8\textwidth]{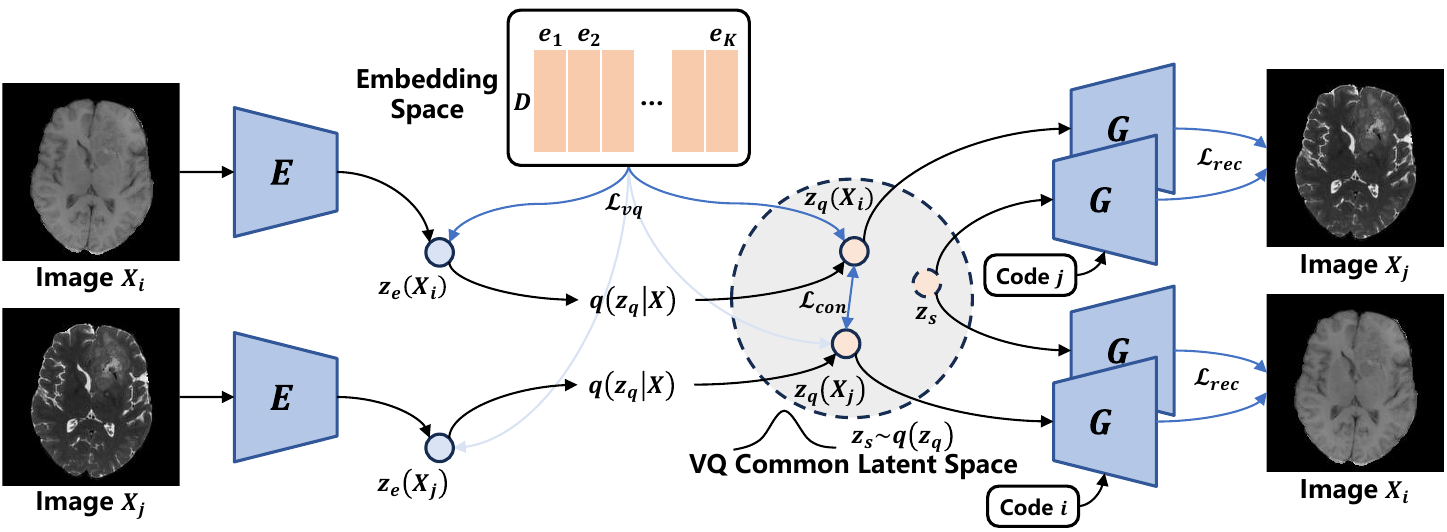}}
    \caption{Overview of the proposed \model~framework.} \label{fig:framework}
\end{figure}
Inheriting the advantages of discrete representations and dynamic models, we propose~\model~to establish the VQC latent space for multi-sequence MRI. As shown in Fig.~\ref{fig:framework}, continuous latent space $z_e$ and corresponding discrete latent space $z_q$ are extracted from the input images. By statistics on $z_q$, we can estimate a VQC latent space containing sampling points $z_s$ that can reconstruct images of different sequences through the dynamic decoder $\mathbf{G}$.
\subsubsection{Uncertainty Estimation}
It is challenging to strictly constrain multi-sequence MRI equal in latent space because one sequence involves specific information that other sequences lack~\cite{sharma2019missing,han2024synthesis}. To tolerate the sequence-specific attributes, we depict the probabilistic scope of $z_q$ among different sequences by considering the uncertainty of the latent space. We propose a simple non-parametric method using the statistics of $z_q$ for uncertainty estimation.
\begin{equation}
    \label{eq:uncertainty_estimate}
    \begin{aligned}
    \mu_q(\mathcal{X}) &= \frac{1}{\sum_{i=1}^N{f_i}}\sum_{i}^{f_i\neq0}{z_q(X_i)} \\
    \sigma^2_q(\mathcal{X}) &= \frac{1}{\sum_{i=1}^N{f_i}-1}\sum_{i}^{f_i\neq0}{(z_q(X_i)-\mu_q(\mathcal{X}))^2}
    \end{aligned}
\end{equation}
\subsubsection{VQC Latent Space}
After obtaining the uncertainty estimation, we can establish a Gaussian distribution for probabilistic statistics. To utilize randomness in further modeling the uncertainty, we use random sampling to draw the VQC latent space from the corresponding distribution randomly.
\begin{equation}
    \label{eq:zs}
    z_s = \mu_q(\mathcal{X})+\epsilon\cdot\sigma^2_q(\mathcal{X}), \quad \epsilon\sim\mathcal{N}(0,1)
\end{equation}
Here, we use the re-parameterization trick to make the sampling operation differentiable, and $\epsilon$ follows the standard Gaussian distribution.
\subsection{Loss Function}
\subsubsection{Pixel-Level Reconstruction}
We establish constraints between the generated image $\hat{X}$ and the target image $X$ at the pixel, structural, and perceptual levels,
\begin{equation}
    \label{eq:rec}
    \mathcal{L}_\text{rec}(\hat{X},X)=\lambda_1\cdot\|\hat{X}-X\|_1 + \lambda_2\cdot\mathcal{L}_\text{ssim}(\hat{X},X) + \lambda_3\cdot\mathcal{L}_\text{per}(\hat{X},X)
\end{equation}
where $\|\cdot\|_1$ refers to the $L_1$ loss, $\mathcal{L}_\text{ssim}$ indicates the SSIM loss~\cite{wang2004image}, and $\mathcal{L}_\text{per}$ presents the perceptual loss~\cite{zhang2018unreasonable} based on pre-trained VGG19. $\lambda_1$, $\lambda_2$, and $\lambda_3$ are weight terms and are experimentally set to be 10, 1, and 0.1.
\subsubsection{Latent Space Consistency}
We ensure that $z_q$ of sequences are close to narrowing the scope of VQC latent space. For two $z_q$ ($z_1$ and $z_2$), we define a consistency loss composed with MSE and contrastive learning loss~\cite{park2020contrastive,hu2022qs}.
\begin{equation}
    \label{eq:con}
    \begin{aligned}
    \mathcal{L}_\text{con}(z_1, z_2)&=\left\|\text{sg}\left[z_1\right]-z_2\right\|_2^2+\left\|\text{sg}\left[z_2\right]-z_1\right\|_2^2 \\
    &-\sum_{p\in M}\log{\frac{\exp{(z_1^{(p)}\cdot z_2^{(p)}/\tau)}}{\sum_{q\in M}{\exp{(z_1^{(p)}\cdot z_2^{(q)}/\tau)}}}\cdot\frac{\exp{(z_2^{(p)}\cdot z_1^{(p)}/\tau)}}{\sum_{q\in M}{\exp{(z_2^{(p)}\cdot z_1^{(q)}/\tau)}}}}
    \end{aligned}
\end{equation}
where $p$ and $q$ are features traversed from pixels in foreground of $z_1$ and $z_2$, $\tau=0.07$ refers to the scalar temperature parameter.
\subsubsection{Total Loss}
We formulate the total loss function using intermediate sequences without adversarial learning.
\begin{equation}
    \label{eq:total}
    \begin{aligned}
    \mathcal{L}_\text{total} & =
    \sum_{i}^{f_i\neq0}\sum_{j}^{f_j\neq0}{\mathcal{L}_\text{rec}(\hat{X}_{i\rightarrow j},X_j)} + \sum_{i}^{f_i\neq0}{\mathcal{L}_\text{rec}(\hat{X}_{s\rightarrow i},X_i)} \\
    + & \lambda_\text{con}\cdot\sum_{i}^{f_i\neq0}\sum_{j}^{f_j\neq0}{\mathcal{L}_\text{con}(z_q(X_i),z_q(X_j))} +\lambda_\text{vq}\cdot\sum_{i}^{f_i\neq0}{\mathcal{L}_\text{vq}(z_e(X_i),z_q(X_i))}
    \end{aligned}
\end{equation}
where $\hat{X}_{i\rightarrow j}=\mathbf{G}(z_q(\mathbf{E}(X_i)),c_j)$ and $\hat{X}_{s\rightarrow j}=\mathbf{G}(z_s,c_j)$ are images generated from $z_q$ and $z_s$, respectively. $\lambda_\text{con}$ and $\lambda_\text{vq}$ are both experimentally set to be 10.
\subsection{Random Domain Augmentation}
We use random domain augmentation for input images during training to further improve the stability of~\model~and the anti-interference ability of VQC latent space. The domain augmentation process has three aspects: (1) simple intensity transformation $\mathcal{T}$ (\textit{e.g.}, gamma transformation, random noise, and bias field); (2) cross-sequence translation with one-hot codes $c_{r}$; and (3) random domain translation with random target codes $c_{r}\sim\mathcal{U}(0,1)$. The latter two augmentation methods allow us to generate an augmented image $X_{\text{aug}}=\mathbf{G}(z_q(X_i),c_r)$ from the input image $X_i$, and the first method makes $X_{\text{aug}}=\mathcal{T}(X_i)$. During training, we will randomly replace the input image $X_i$ with one of $X_\text{aug}$.
\begin{figure}[b]
\centering
    \begin{minipage}{0.1\linewidth}
        \centerline{(a)}
    \end{minipage}
    \begin{minipage}{0.28\linewidth}
        \centerline{\includegraphics[width=\textwidth]{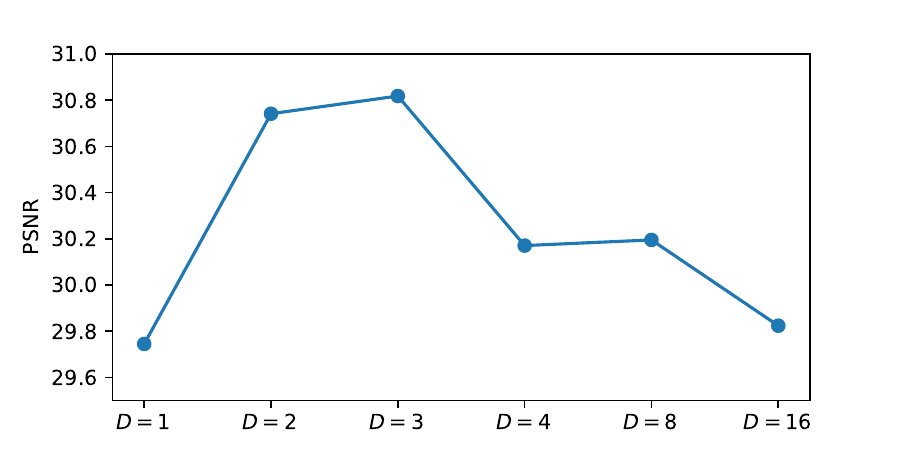}}
    \end{minipage}
    \begin{minipage}{0.28\linewidth}
        \centerline{\includegraphics[width=\textwidth]{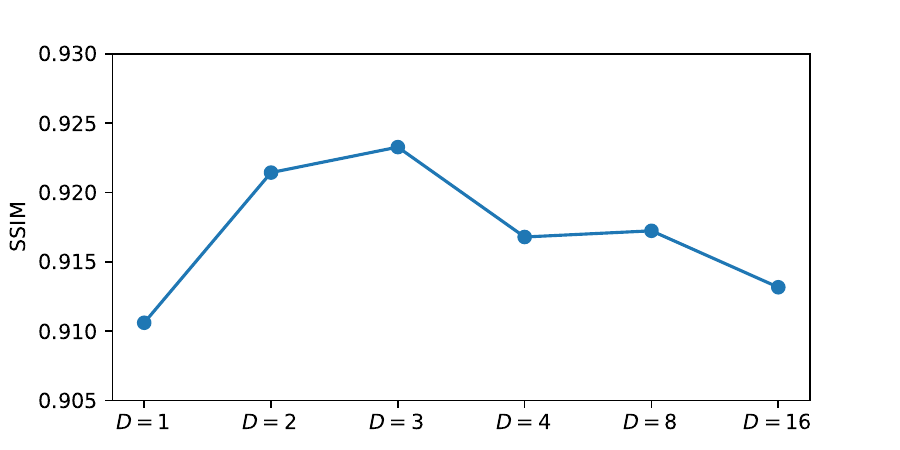}}
    \end{minipage}
    \begin{minipage}{0.28\linewidth}
        \centerline{\includegraphics[width=\textwidth]{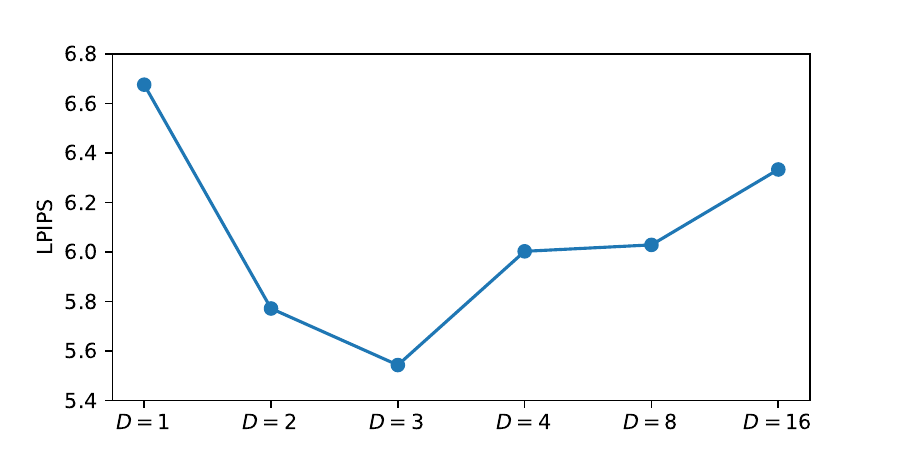}}
    \end{minipage}


    \begin{minipage}{0.1\linewidth}
        \centerline{(b)}
    \end{minipage}
    \begin{minipage}{0.28\linewidth}
        \centerline{\includegraphics[width=\textwidth]{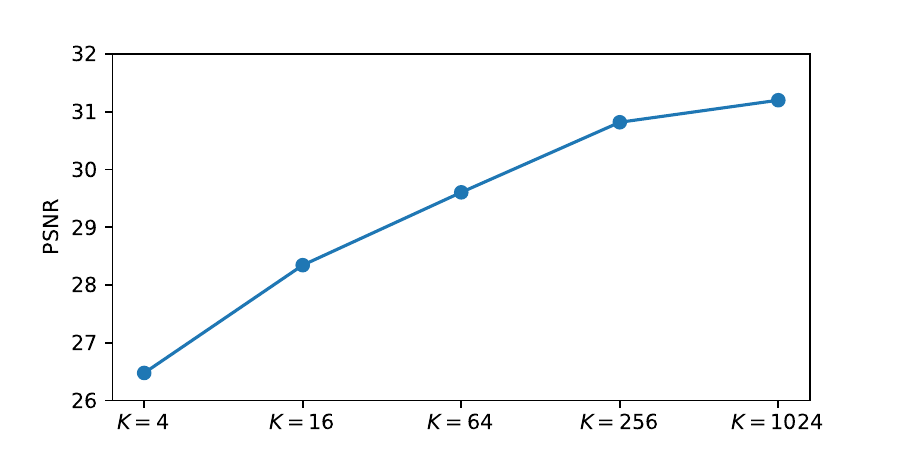}}
    \end{minipage}
    \begin{minipage}{0.28\linewidth}
        \centerline{\includegraphics[width=\textwidth]{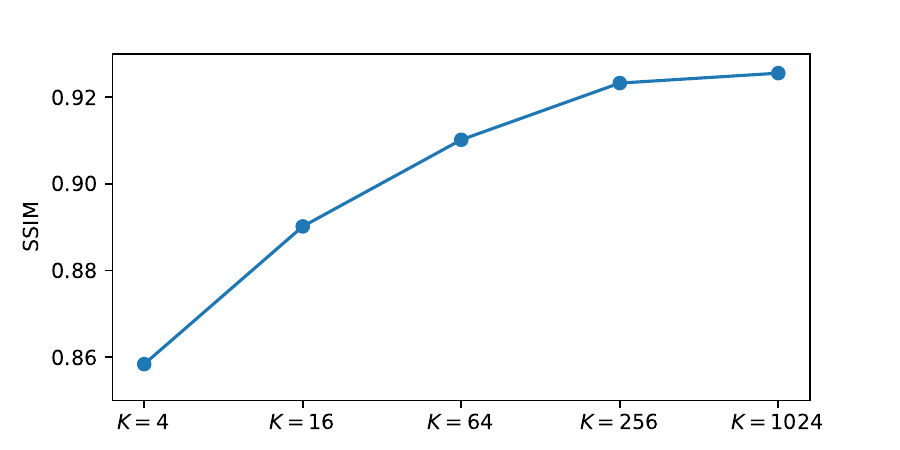}}
    \end{minipage}
    \begin{minipage}{0.28\linewidth}
        \centerline{\includegraphics[width=\textwidth]{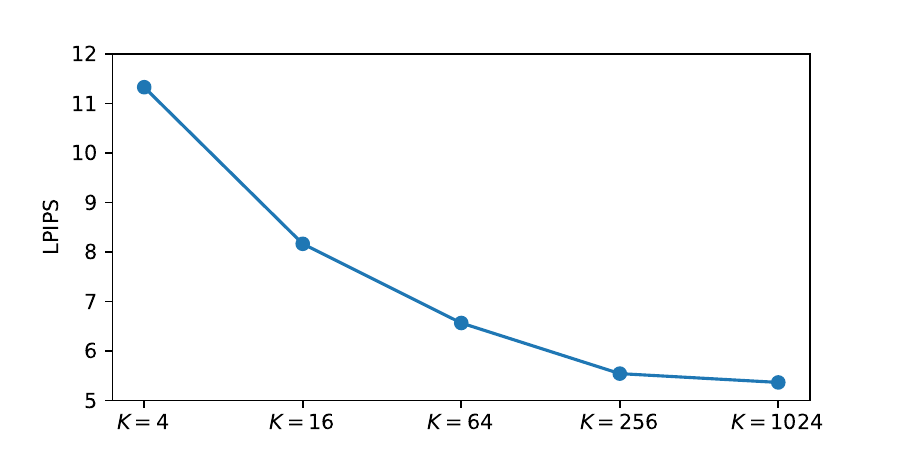}}
    \end{minipage}

    
    \caption{Synthesis performance of~\model~with different latent dimensions ($D$) and embedding dimensions ($K$). (a) Synthesis performance with different latent dimensions ($K=256$); (b) Synthesis performance with different embedding dimensions ($D=3$).} \label{fig:dim}
\end{figure}
\section{Experiments}
\subsection{Experimental Settings}
\subsubsection{Dataset and Evaluation Metrics}\label{sec:dataset}
We utilize brain MRI images from the Brain Tumor Segmentation 2021 (BraTS2021) dataset~\cite{menze2014multimodal,bakas2017advancing,baid2021rsna}, comprising 1,251 subjects with four aligned sequences: T1, T1Gd, T2, and Flair. From this dataset, we allocated 830 subjects for training, 93 for validation, and 328 for testing. To simulate clinical settings with missing sequences, we divided the training set into three subsets, which contained paired sequences between (T1, T1Gd), (T1Gd, T2), and (T2, Flair), respectively. It can be simulated that there is no paired sample between T1 and Flair under this setting, but there are two partially paired intermediate sequences, T1Gd and T2. All images undergo intensity normalization to a range of [0, 1] and are subsequently centrally cropped to dimensions of 128$\times$192$\times$192.
Synthesis performance is evaluated using metrics including peak signal-to-noise ratio (PSNR), structural similarity index measure (SSIM), and learned perceptual image patch similarity (LPIPS).
\subsubsection{Implementation Details}
We implemented the models using PyTorch and trained them on the NVIDIA GeForce RTX 3090 Ti GPU. The architecture of $\mathbf{E}$ and $\mathbf{G}$ is the same as Seq2Seq~\cite{han2024synthesis}. The proposed~\model~is trained using the AdamW optimizer, with an initial learning rate of $10^{-4}$ and a batch size of 1 for 1,000,000 steps. All comparative experiments use domain augmentation, at least with simple intensity transformation $\mathcal{T}$, to ensure a fair comparison. $\mathcal{T}$ involves applying random gamma transformation with $\gamma\sim\mathcal{U}(0.95,1.05)$, random Gaussian noise with $\sigma\sim\mathcal{U}(0,0.1)$, and random bias field with scale of 0.2 and degree of intensity inhomogeneity $\alpha\sim\mathcal{U}(0,2)$.
\begin{table}[t]
    \caption{The quantitative results of translating T1 to T1Gd, T2, and Flair with a single step or multiple steps. The best result is in bold, and the second best is underlined.}
    \label{tab:step}
    \setlength{\tabcolsep}{2pt}
    \begin{adjustbox}{width=\columnwidth,center}
    \begin{tabular}{llccccccccccc}
        \hline\hline
        \multirow{2}*{Step} & \multirow{2}*{Method}
            & \multicolumn{3}{c}{T1$\rightarrow$T1Gd} && \multicolumn{3}{c}{T1$\rightarrow$(T1Gd)$\rightarrow$T2} && \multicolumn{3}{c}{T1$\rightarrow$(T1Gd$\rightarrow$T2)$\rightarrow$Flair} \\
            \cline{3-5}\cline{7-9}\cline{11-13}
            && PSNR$\uparrow$ & SSIM$\uparrow$ & LPIPS$\downarrow$ && PSNR$\uparrow$ & SSIM$\uparrow$ & LPIPS$\downarrow$ && PSNR$\uparrow$ & SSIM$\uparrow$ & LPIPS$\downarrow$ \\
        \hline
        \multirow{8}*{Single}
            & MM-GAN~\cite{sharma2019missing} & 26.5$\pm$2.0 & 0.871$\pm$0.044 & 11.4$\pm$4.2 && 22.2$\pm$1.0 & 0.779$\pm$0.028 & 20.5$\pm$3.4 && 21.7$\pm$1.2 & 0.791$\pm$0.039 & 17.8$\pm$3.1 \\
            & ResViT~\cite{dalmaz2022resvit} & 26.4$\pm$2.0 & 0.872$\pm$0.040 & 11.6$\pm$4.1 && 21.8$\pm$0.9 & 0.774$\pm$0.035 & 16.1$\pm$3.2 && 20.9$\pm$0.9 & 0.705$\pm$0.035 & 20.0$\pm$3.4 \\
            & Jiang~\textit{et al.}~\cite{jiang2020unified} & 26.7$\pm$2.7 & 0.874$\pm$0.044 & 10.4$\pm$4.1 && 23.7$\pm$2.2 & 0.833$\pm$0.039 & 12.3$\pm$3.9 && 23.5$\pm$2.4 & 0.796$\pm$0.054 & 11.9$\pm$3.6 \\
            \cline{2-13}
            & Seq2Seq~\cite{han2024synthesis} & 26.9$\pm$2.2 & 0.876$\pm$0.040 & \underline{9.68$\pm$3.74} && 26.5$\pm$1.9 & 0.884$\pm$0.039 & 8.25$\pm$3.32 && 24.3$\pm$2.3 & 0.811$\pm$0.047 & 11.2$\pm$3.5 \\
            & \quad +VQ & 27.0$\pm$2.2 & 0.875$\pm$0.040 & 9.79$\pm$3.67 && 26.7$\pm$1.9 & 0.885$\pm$0.038 & 8.05$\pm$3.22 && 24.6$\pm$2.3 & 0.817$\pm$0.043 & 11.0$\pm$3.5 \\
            & \quad +VQ+$\mathcal{L}_\text{con}$ & \underline{27.1$\pm$2.1} & \underline{0.876$\pm$0.039} & \textbf{9.67$\pm$3.61} && 26.8$\pm$1.9 & \underline{0.886$\pm$0.038} & 7.85$\pm$3.30 && 24.7$\pm$2.2 & 0.823$\pm$0.040 & 10.8$\pm$3.4 \\
            & \model & \textbf{27.1$\pm$2.1} & \textbf{0.876$\pm$0.039} & 9.78$\pm$3.49 && \textbf{27.1$\pm$1.9} & \textbf{0.890$\pm$0.036} & \underline{7.63$\pm$3.07} && 25.7$\pm$1.9 & \textbf{0.847$\pm$0.033} & 9.86$\pm$3.05 \\
            & \quad w/o Aug & 27.0$\pm$2.0 & 0.875$\pm$0.038 & 9.71$\pm$3.41 && 26.8$\pm$1.8 & 0.883$\pm$0.036 & 8.23$\pm$3.17 && 24.1$\pm$2.2 & 0.805$\pm$0.050 & 11.4$\pm$3.5 \\
        \hline
        \multirow{8}*{Multiple}
            & MM-GAN~\cite{sharma2019missing} & - & - & - && 25.4$\pm$1.7 & 0.866$\pm$0.037 & 10.9$\pm$3.4 && 24.9$\pm$1.3 & 0.826$\pm$0.032 & 14.2$\pm$3.4 \\
            & ResViT~\cite{dalmaz2022resvit} & - & - & - && 25.7$\pm$1.7 & 0.861$\pm$0.032 & 10.7$\pm$3.2 && 24.9$\pm$1.1 & 0.826$\pm$0.037 & 14.5$\pm$4.2 \\
            & Jiang~\textit{et al.}~\cite{jiang2020unified} & - & - & - && 26.0$\pm$1.9 & 0.874$\pm$0.037 & 9.70$\pm$3.16 && 25.3$\pm$1.6 & 0.835$\pm$0.030 & 10.5$\pm$3.8 \\
            \cline{2-13}
            & Seq2Seq~\cite{han2024synthesis} & - & - & - && 26.4$\pm$1.8 & 0.883$\pm$0.037 & 7.94$\pm$3.03 && 25.5$\pm$1.5 & 0.843$\pm$0.029 & 9.83$\pm$2.84 \\
            & \quad +VQ & - & - & - && 26.4$\pm$1.8 & 0.878$\pm$0.037 & 8.15$\pm$2.98 && 25.5$\pm$1.5 & 0.839$\pm$0.028 & 9.83$\pm$2.65 \\
            & \quad +VQ+$\mathcal{L}_\text{con}$ & - & - & - && 26.6$\pm$1.8 & 0.881$\pm$0.036 & 7.74$\pm$2.89 && 25.7$\pm$1.6 & 0.843$\pm$0.028 & \underline{9.63$\pm$2.67} \\
            & \model & - & - & - && \underline{26.8$\pm$1.8} & 0.884$\pm$0.035 & \textbf{7.63$\pm$2.76} && \textbf{25.9$\pm$1.5} & \underline{0.846$\pm$0.028} & \textbf{9.47$\pm$2.56} \\
            & \quad w/o Aug & - & - & - && 26.6$\pm$1.7 & 0.877$\pm$0.034 & 8.23$\pm$2.81 && \underline{25.7$\pm$1.4} & 0.840$\pm$0.028 & 11.1$\pm$2.7 \\
        \hline\hline
    \end{tabular}%
    \end{adjustbox}
\end{table}
\begin{figure}[t]
\centering
    \begin{minipage}{0.17\linewidth}
        \centerline{}
    \end{minipage}
    \begin{minipage}{0.145\linewidth}
        \centerline{MM-GAN}
    \end{minipage}
    \begin{minipage}{0.145\linewidth}
        \centerline{Jiang~\textit{et al.}}
    \end{minipage}
    \begin{minipage}{0.145\linewidth}
        \centerline{Seq2Seq}
    \end{minipage}
    \begin{minipage}{0.145\linewidth}
        \centerline{\model}
    \end{minipage}
    \begin{minipage}{0.145\linewidth}
        \centerline{Target}
    \end{minipage}

    \begin{minipage}{0.17\linewidth}
        \centerline{T1Gd}
    \end{minipage}
    \begin{minipage}{0.145\linewidth}
        \centerline{\includegraphics[width=\textwidth]{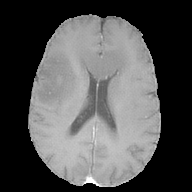}}
    \end{minipage}
    \begin{minipage}{0.145\linewidth}
        \centerline{\includegraphics[width=\textwidth]{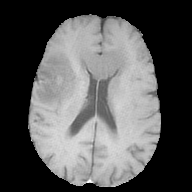}}
    \end{minipage}
    \begin{minipage}{0.145\linewidth}
        \centerline{\includegraphics[width=\textwidth]{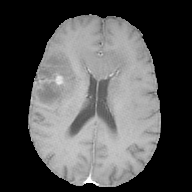}}
    \end{minipage}
    \begin{minipage}{0.145\linewidth}
        \centerline{\includegraphics[width=\textwidth]{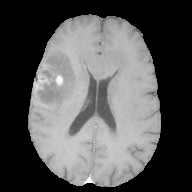}}
    \end{minipage}
    \begin{minipage}{0.145\linewidth}
        \centerline{\includegraphics[width=\textwidth]{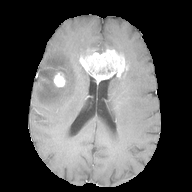}}
    \end{minipage}

    \begin{minipage}{0.17\linewidth}
        \centerline{T2}
    \end{minipage}
    \begin{minipage}{0.145\linewidth}
        \centerline{\includegraphics[width=\textwidth]{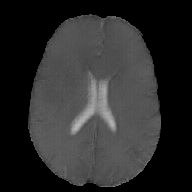}}
    \end{minipage}
    \begin{minipage}{0.145\linewidth}
        \centerline{\includegraphics[width=\textwidth]{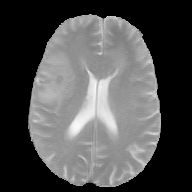}}
    \end{minipage}
    \begin{minipage}{0.145\linewidth}
        \centerline{\includegraphics[width=\textwidth]{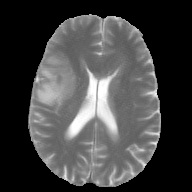}}
    \end{minipage}
    \begin{minipage}{0.145\linewidth}
        \centerline{\includegraphics[width=\textwidth]{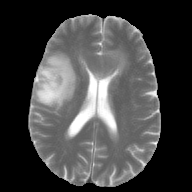}}
    \end{minipage}
    \begin{minipage}{0.145\linewidth}
        \centerline{\includegraphics[width=\textwidth]{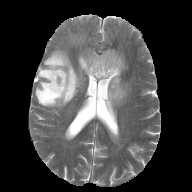}}
    \end{minipage}

    \begin{minipage}{0.17\linewidth}
        \centerline{Flair}
    \end{minipage}
    \begin{minipage}{0.145\linewidth}
        \centerline{\includegraphics[width=\textwidth]{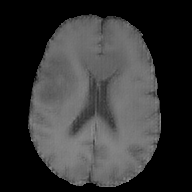}}
    \end{minipage}
    \begin{minipage}{0.145\linewidth}
        \centerline{\includegraphics[width=\textwidth]{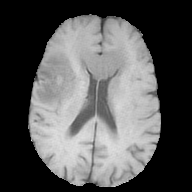}}
    \end{minipage}
    \begin{minipage}{0.145\linewidth}
        \centerline{\includegraphics[width=\textwidth]{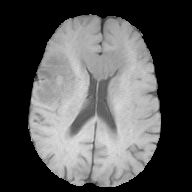}}
    \end{minipage}
    \begin{minipage}{0.145\linewidth}
        \centerline{\includegraphics[width=\textwidth]{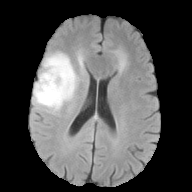}}
    \end{minipage}
    \begin{minipage}{0.145\linewidth}
        \centerline{\includegraphics[width=\textwidth]{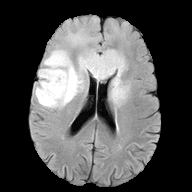}}
    \end{minipage}

    \caption{Visualization of translating T1 to T1Gd, T2, and Flair with a single step.} \label{fig:step}
\end{figure}
\subsection{Experimental Results}
\subsubsection{Latent and Embedding Dimension}
Referring to Sec.~\ref{sec:vqvae}, the latent dimension $D$ represents the dimension of the compressed feature. The smaller $D$ is, the greater the degree of compression. The embedding dimension $K$ indicates the number of discrete vectors (clustering) in the codebook. The larger $K$ is, the better a discrete vector fits the continuous features.
We train~\model~using the training sets with complete sequences to explore the optimal $D$ and $K$ before other experiments.
As shown in Fig.~\ref{fig:dim}, when $K=256$, the proposed model performs the best when $D=3$. Additionally, when $D=3$, the performance of the model continues to improve as $K$ increases, but the rate of improvement slows down after $K>256$. Thus, we set $D=3$ and $K=256$ in this work.
\begin{table}[t]
    \caption{The quantitative results for comparisons of reconstructing images based on noise and bias field data. The best result is in bold, and the second best is underlined.}
    \label{tab:anti}
    \setlength{\tabcolsep}{4pt}
    \begin{adjustbox}{width=\columnwidth,center}
    \begin{tabular}{lccccccc}
        \hline\hline
        \multirow{2}*{Method}
            & \multicolumn{3}{c}{Noise} && \multicolumn{3}{c}{Bias Field} \\
            \cline{2-4}\cline{6-8}
            & PSNR$\uparrow$ & SSIM$\uparrow$ & LPIPS$\downarrow$ && PSNR$\uparrow$ & SSIM$\uparrow$ & LPIPS$\downarrow$ \\
        \hline
        MM-GAN~\cite{sharma2019missing} & 29.0$\pm$0.5 & 0.860$\pm$0.028 & 20.6$\pm$5.1 && 22.1$\pm$0.7 & 0.928$\pm$0.015 & 5.56$\pm$1.43 \\
        ResViT~\cite{dalmaz2022resvit} & 28.6$\pm$2.3 & 0.851$\pm$0.023 & 19.8$\pm$8.6 && 20.0$\pm$1.7 & 0.914$\pm$0.016 & 6.37$\pm$1.88 \\
        Jiang~\textit{et al.}~\cite{jiang2020unified} & 30.1$\pm$1.8 & \underline{0.895$\pm$0.021} & 10.3$\pm$3.1 && 21.2$\pm$0.8 & 0.924$\pm$0.019 & 5.98$\pm$1.18 \\
        \hline
        Seq2Seq~\cite{han2024synthesis} & 28.2$\pm$2.6 & 0.861$\pm$0.024 & 14.6$\pm$3.9 && 22.0$\pm$1.2 & 0.927$\pm$0.019 & 5.49$\pm$1.23 \\
        \quad +VQ & 29.0$\pm$2.9 & 0.877$\pm$0.027 & 10.7$\pm$3.1 && 22.4$\pm$1.2 & 0.928$\pm$0.018 & 5.14$\pm$1.31 \\
        \quad +VQ+$\mathcal{L}_\text{con}$ & \underline{30.3$\pm$1.5} & 0.891$\pm$0.017 & \underline{9.38$\pm$2.73} && 22.6$\pm$1.4 & \textbf{0.930$\pm$0.019} & \textbf{5.09$\pm$1.27} \\
        \model & \textbf{30.3$\pm$1.5} & \textbf{0.902$\pm$0.016} & \textbf{7.02$\pm$1.73} && \textbf{26.1$\pm$2.6} & \underline{0.930$\pm$0.020} & \underline{5.09$\pm$1.51} \\
        \quad w/o Aug & 29.2$\pm$2.5 & 0.864$\pm$0.044 & 10.7$\pm$2.9 && 22.6$\pm$1.6 & 0.916$\pm$0.019 & 6.23$\pm$1.46 \\
        \hline\hline
    \end{tabular}%
    \end{adjustbox}
\end{table}
\begin{figure}[t]
\centering
    \begin{minipage}{0.17\linewidth}
        \centerline{}
    \end{minipage}
    \begin{minipage}{0.145\linewidth}
        \centerline{Input}
    \end{minipage}
    \begin{minipage}{0.145\linewidth}
        \centerline{Jiang~\textit{et al.}}
    \end{minipage}
    \begin{minipage}{0.145\linewidth}
        \centerline{Seq2Seq}
    \end{minipage}
    \begin{minipage}{0.145\linewidth}
        \centerline{\model}
    \end{minipage}
    \begin{minipage}{0.145\linewidth}
        \centerline{Target}
    \end{minipage}

    \begin{minipage}{0.17\linewidth}
        \centerline{Artifacts}
    \end{minipage}
    \begin{minipage}{0.145\linewidth}
        \centerline{\includegraphics[width=\textwidth]{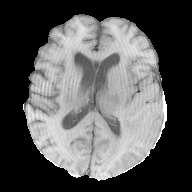}}
    \end{minipage}
    \begin{minipage}{0.145\linewidth}
        \centerline{\includegraphics[width=\textwidth]{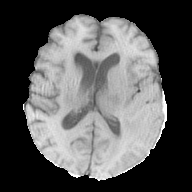}}
    \end{minipage}
    \begin{minipage}{0.145\linewidth}
        \centerline{\includegraphics[width=\textwidth]{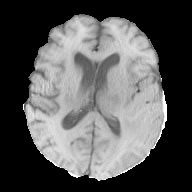}}
    \end{minipage}
    \begin{minipage}{0.145\linewidth}
        \centerline{\includegraphics[width=\textwidth]{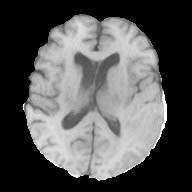}}
    \end{minipage}
    \begin{minipage}{0.145\linewidth}
        \centerline{N/A}
    \end{minipage}

    \begin{minipage}{0.17\linewidth}
        \centerline{Noise}
    \end{minipage}
    \begin{minipage}{0.145\linewidth}
        \centerline{\includegraphics[width=\textwidth]{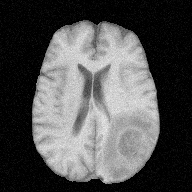}}
    \end{minipage}
    \begin{minipage}{0.145\linewidth}
        \centerline{\includegraphics[width=\textwidth]{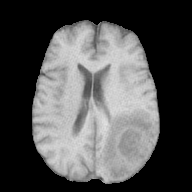}}
    \end{minipage}
    \begin{minipage}{0.145\linewidth}
        \centerline{\includegraphics[width=\textwidth]{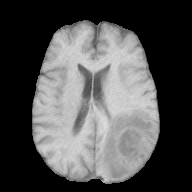}}
    \end{minipage}
    \begin{minipage}{0.145\linewidth}
        \centerline{\includegraphics[width=\textwidth]{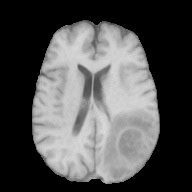}}
    \end{minipage}
    \begin{minipage}{0.145\linewidth}
        \centerline{\includegraphics[width=\textwidth]{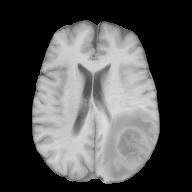}}
    \end{minipage}

    \begin{minipage}{0.17\linewidth}
        \centerline{Bias Field}
    \end{minipage}
    \begin{minipage}{0.145\linewidth}
        \centerline{\includegraphics[width=\textwidth]{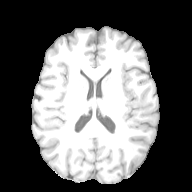}}
    \end{minipage}
    \begin{minipage}{0.145\linewidth}
        \centerline{\includegraphics[width=\textwidth]{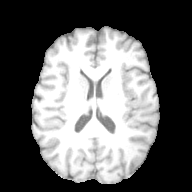}}
    \end{minipage}
    \begin{minipage}{0.145\linewidth}
        \centerline{\includegraphics[width=\textwidth]{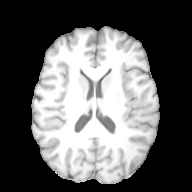}}
    \end{minipage}
    \begin{minipage}{0.145\linewidth}
        \centerline{\includegraphics[width=\textwidth]{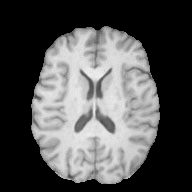}}
    \end{minipage}
    \begin{minipage}{0.145\linewidth}
        \centerline{\includegraphics[width=\textwidth]{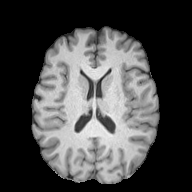}}
    \end{minipage}

    \caption{Visualization of reconstruction from input images with artifacts, noise, and bias field. Artifacts exist in the original images, therefore, the target image is unavailable.} \label{fig:anti}
\end{figure}
\subsubsection{Latent Space Consistency}
To evaluate the effectiveness of the proposed VQC latent space for unsupervised cross-sequence generation, we compared~\model~with other methods such as MM-GAN~\cite{sharma2019missing}, ResViT~\cite{dalmaz2022resvit}, Jiang~\textit{et al.}~\cite{jiang2020unified}, and Seq2Seq~\cite{han2024synthesis}. Additionally, we compared the three components of our method, which include VQ embedding, VQ with $\mathcal{L}_\text{con}$, and domain augmentation. There are two ways to implement a source$\rightarrow$target generation: (1) generate the target directly from the source (single-step), and (2) first generate an intermediate sequence from the source and then generate the target (multi-step). Table~\ref{tab:step} and Fig.~\ref{fig:step} illustrate the synthesis performance of comparisons on translating T1$\rightarrow$T1Gd, T1$\rightarrow$T2, and T1$\rightarrow$Flair. Note that, due to the settings of paired samples in the training set, the multi-step generation between T1 and T2 requires two steps, and between T1 and Flair requires three steps.
As shown in Table~\ref{tab:step}, the comparison method achieves similar performance for the T1$\rightarrow$T1Gd generation task with paired samples. However, when it comes to unpaired T1$\rightarrow$T2 and T1$\rightarrow$Flair generation tasks, the performance of the comparison method decreases sharply when performing single-step generation compared to multi-step generation. In contrast, the proposed~\model~shows only a minor performance penalty on T1$\rightarrow$Flair task and improves on T1$\rightarrow$T2 task. This shows that multi-step generation will lead to information loss and error accumulation, and our~\model~can alleviate this problem through single-step generation.
\subsubsection{Anti-Interference}
The proposed VQC latent space also has the anti-interference ability. We add fixed Gaussian noise and bias fields to the input image and reconstruct the input image using the comparisons. As shown in Table~\ref{tab:anti}, the proposed method can effectively prevent the interference of noise and bias fields to reconstruct the original image. Fig.~\ref{fig:anti} shows the visualization results of the reconstruction, in which we found that the proposed model can also remove artifacts in images.
\subsubsection{Compression and Representation}
\begin{table}[t]
    \caption{The quantitative one-shot segmentation results for using latent space from comparisons. The best result is in bold. ET: enhanced tumor, TC: tumor core, WT: whole tumor.}
    \label{tab:seg}
    \setlength{\tabcolsep}{4pt}
    \begin{adjustbox}{width=\columnwidth,center}
    \begin{tabular}{lccccccc}
        \hline\hline
        \multirow{2}*{Method}
            & \multicolumn{3}{c}{DSC$\uparrow$} && \multicolumn{3}{c}{ASSD$\downarrow$} \\
            \cline{2-4}\cline{6-8}
            & ET & TC & WT && ET & TC & WT \\
        \hline
        nnU-Net~\cite{isensee2021nnu} & 0.481$\pm$0.298 & 0.457$\pm$0.308 & 0.463$\pm$0.231 && 12.7$\pm$11.9 & 13.7$\pm$11.2 & 14.2$\pm$7.1 \\
        Jiang~\textit{et al.}~\cite{jiang2020unified} & 0.193$\pm$0.276 & 0.175$\pm$0.272 & 0.328$\pm$0.272 && 24.3$\pm$10.3 & 23.6$\pm$10.6 & 15.9$\pm$8.8 \\
        Seq2Seq~\cite{han2024synthesis} & 0.276$\pm$0.234 & 0.274$\pm$0.224 & 0.386$\pm$0.206 && 19.3$\pm$9.4 & 19.4$\pm$8.4 & 15.6$\pm$5.4 \\
        \model & \textbf{0.557$\pm$0.275} & \textbf{0.532$\pm$0.301} & \textbf{0.638$\pm$0.201} && \textbf{6.46$\pm$9.74} & \textbf{8.18$\pm$9.97} & \textbf{9.36$\pm$5.96} \\
        \hline\hline
    \end{tabular}%
    \end{adjustbox}
\end{table}
\begin{figure}[t]
\centering
    \begin{minipage}{0.37\linewidth}
        \centerline{\includegraphics[width=\textwidth]{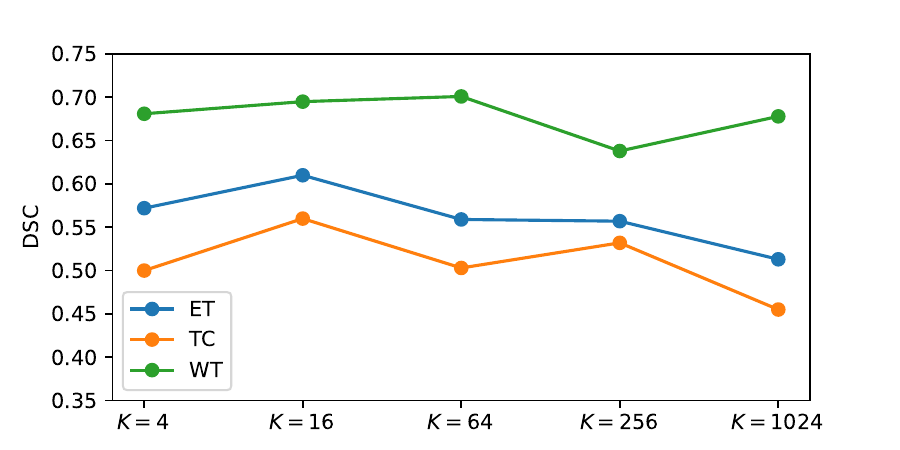}}
    \end{minipage}
    \begin{minipage}{0.37\linewidth}
        \centerline{\includegraphics[width=\textwidth]{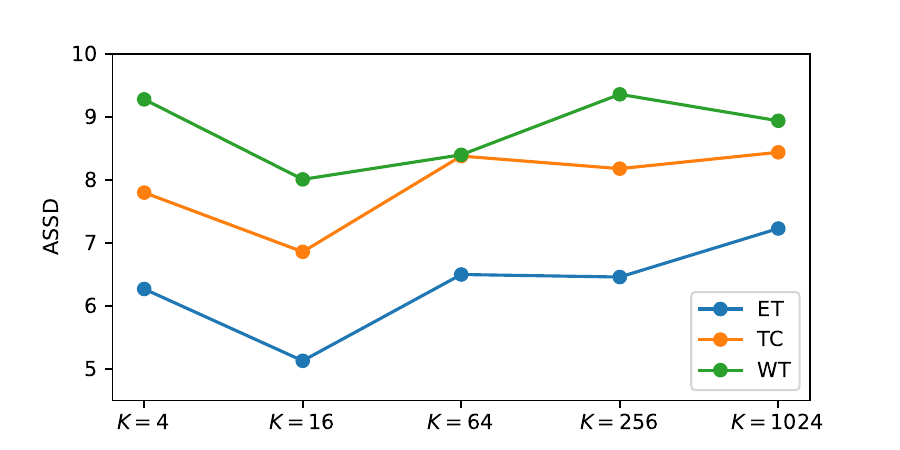}}
    \end{minipage}
    
    \caption{One-shot segmentation performance of \model~with different embedding dimensions ($K$).} \label{fig:seg}
\end{figure}
The proposed VQC latent space showcases strong representation ability, indicating the potential of one-shot segmentation. To demonstrate this, we train the nnU-Net model based on the VQC latent space for brain tumor segmentation. For this purpose, we only use one subject containing all sequences for training. As shown in Table~\ref{tab:seg}, the segmentation model trained based on the VQC latent space outperforms the model trained using only images. Furthermore, Fig.~\ref{fig:seg} shows that fewer VQ embedding dimensions $K=16$ contribute towards the clustering of image semantics, which improves the segmentation performance.
\section{Conclusion}
In this work, we introduce a network for estimating the distribution of VQC latent space, which inherits the advantage of discrete representations and dynamic models. Experimental results based on BraTS2021 demonstrate that this latent space contributes to cross-sequence generation without adversarial learning and has substantial anti-interference and representation ability.

\begin{credits}
\subsubsection{\ackname} Luyi Han was funded by Chinese Scholarship Council (CSC) scholarship.
\end{credits}
%
%
%
%
\bibliographystyle{splncs04}
\bibliography{Paper-0936}
\end{document}